\documentclass[a4,11pt]{article}
\usepackage{epsfig}

\usepackage{amssymb,epsfig,graphicx,epsf}

\usepackage{amsfonts}
\usepackage{amsmath}
\usepackage{amssymb}
\newcommand{\be}{\begin{equation}}
\newcommand{\ee}{\end{equation}}

\newcommand{\ba}{\begin{eqnarray}}
\newcommand{\ea}{\end{eqnarray}}

\begin{document}

\title{The emergence of geometry: a two-dimensional toy model}

\author{Jorge Alfaro\\
Pontificia Universidad Cat\'olica de Chile, Av. Vicu\~na Mackenna 4860, Santiago, Chile,\\
\\
Dom\`enec Espriu\footnote{On leave of absence from ICCUB and DECM, 
Universitat de Barcelona}\\ 
CERN, 1211 Geneva 23, Switzerland,\\
\\ 
and \\
\\
Daniel Puigdom\`enech\\
Departament d'Estructura i Constituents de la Mat\`eria  and\\ 
Institut de Ci\`encies del Cosmos (ICCUB)\\ Universitat de Barcelona\\
Mart\'\i ~i Franqu\`es, 1, 08028 Barcelona, Spain.}

\date{}

\maketitle

\begin{abstract}
We review the similarities between the effective chiral lagrangrian, relevant for
low-energy strong interactions, and the Einstein-Hilbert action. We use these
analogies to suggest a specific mechanism whereby
gravitons would emerge as Goldstone bosons of a global  
$SO(D)\times GL(D)$ symmetry broken down to $SO(D)$ by fermion condensation.      
We propose a two-dimensional toy 
model where a dynamical zwei-bein is generated from a topological theory without 
any pre-existing metric structure, the space being endowed only with an affine connection.
A metric appears only after the symmetry breaking; thus the notion
of distance is an induced effective one.
In spite of several non-standard features this simple toy model appears 
to be renormalizable and at long distances is described by an effective lagrangian that 
corresponds to that of two-dimensional gravity (Liouville theory). The induced cosmological
constant is related to the dynamical mass $M$ acquired by the fermion fields in the breaking, which
also acts as an infrared regulator. The low-energy expansion is valid for momenta $k >M$, i.e. for
supra-horizon scales. We briefly
discuss a possible implementation of a similar mechanism in four dimensions.

\end{abstract}

\vfill
\noindent
March 2010

\noindent
UB-ECM-FP-12/10

\noindent
ICCUB-10-027

\section{Introduction}
Einstein formulated general relativity in 1915 and ninety-five years later we still
have little or no clue as to the true quantum nature of this theory. 

It is surely correct
to say that string theory is able to provide a consistent perturbative quantum theory of 
gravitation, at the price of a rather radical modification of quantum field theory, including
the acceptance that our world has more than four dimensions. Unfortunately string theory
is not able to select a unique vacuum, in particular it does not shed light at 
present on the fact that we live in a world where $\langle g_{\mu\nu}\rangle \neq 0$.
Other modifications of gravity that include extra dimensions, although extremely interesting
from a conceptual and phenomenological point of view, tipically lack a ultraviolet completion   
and therefore should probably find their ultimate justification in specific compactifications
of string theory (where again the choice of vacuum appears itself). 

Less popular alternatives, but of considerable interest nonetheless, are the 
search for non-trivial ultraviolet
fixed points in gravity (asymptotic safety\cite{weias}) and the notion of induced gravity\cite{adler}. 
The former
approach is the one pursued by exact renormalization-group practitioners\cite{litim} and by lattice 
and numerical techniques such as Lorentzian triangulation analysis\cite{loll}. Induced gravity 
advocates that a possible explanation of the relative weakness of gravity as compared to other interactions
is that it is a residual or induced force, a subproduct of all the rest of matter 
and interaction fields.
With the exception of lattice studies, all these approaches also rely on the introduction
of a metric from the very beginning. On the contrary, lattice analysis only require
some pre-metric input, in particular a notion of causality (hence transport of a time-like vector). 

It has been pointed out several times in the literature that gravitons should perhaps 
be considered as Goldstone bosons of some broken symmetry. This is exactly the point
of view that we adopt in this paper. This idea goes probably back to early papers
by Salam and coworkers\cite{salam}, and Ogievetsky and coworkers\cite{ogi}, if not 
earlier\footnote{We thank Luis Alvarez-Gaum\'e for pointing out to us some of the earlier work on this subject}, 
but a concrete proposal 
has been lacking so far (see however \cite{ru}). By concrete proposal we mean some field theory
that does not contain the graviton field as an elementary degree of freedom. Ideally it
should not even contain the tensor $\eta_{\mu\nu}$ as this already implies
the use of some background metric. Indeed we would like to see the metric degrees
of freedom emerging dynamically, like the pions appear dynamically after chiral 
symmetry breaking in QCD. Furthermore, if possible, we would like the underlying theory to be
in some sense 'simpler' than gravity, in particular it should be renormalizable.
One could then pose questions that are left unanswered in gravity, such as the fate
of black hole singularities and the counting of degrees of freedom.

\section{The low-energy effective action of QCD}

The four-dimensional chiral lagrangian is a non-renormalizable theory describing accurately pion
physics at low energies. It has a long history, with the first formal studies concerning 
renormalizability being due mostly to Weinberg\cite{wein} and later considerably extended by
Gasser and Leutwyler\cite{gl}. The chiral lagrangian contains a (infinite) number
of operators 
\be
{\cal L}= f_\pi^2 {\rm Tr\,} \partial_\mu U \partial^\mu U^\dagger
+ \alpha_1 {\rm Tr\,}\partial_\mu U \partial^\mu U^\dagger \partial_\nu U 
\partial^\nu U^\dagger + \alpha_ 2 {\rm Tr\,} \partial_\mu U 
\partial_\nu U^\dagger \partial^\mu U \partial^\nu U^\dagger + \ldots
\ee
\be 
U\equiv  \exp i \tilde \pi/ f_\pi, \qquad \tilde\pi \equiv \pi^a \tau^a/2, 
\nonumber\ee
organized according to the number of derivatives 
\be 
{\cal L} = {\cal O}(p^2) + {\cal O}(p^4) + {\cal O}(p^6) + ...
\ee
Pions are the Goldstone bosons associated to the (global) symmetry breaking
pattern of QCD
\be
SU(2)_L \times SU(2)_R \to SU(2)_V
\ee
The above lagrangian is the most general one compatible with the symmetries 
of QCD and their breaking.
Locality, symmetry and relevance (in the renormalization group sense)  are the 
guiding principles to construct ${\cal L}$. Renormalizability is not; in fact if
we cut-off the derivarive expansion at a given order the theory 
requires counterterms beyond that order no matter how large the order is.
Note that, although the symmetry has been spontaneously broken, the effective 
lagrangian still has the full symmetry $U\to L U R^\dagger$ with
$L$ and $R$ being $SU(2)$ matrices belonging to the left and right groups, respectively.

The lowest-order, tree level contribution to pion-pion
scattering derived from the previous lagrangian is $\sim {p^2}/{f_\pi^2}$. Simple counting
arguments show that the one-loop chiral corrections are $\sim {p^4}/{(16\pi^2 f_\pi^4)}$. 
Thus the counting 
parameter in the loop (chiral) expansion in 4D is
\be
\frac{p^2}{16\pi^2 f_\pi^2}.
\ee
Each chiral loop gives an additional power of $p^2$.

At each order in perturbation theory the divergences that arise can be 
eliminated by redefining the coefficients in the higher order operators
\be 
\alpha_i\to \alpha_i + \frac{c_i}{\epsilon} 
\ee
In addition to the pure pole in $\epsilon$, 
logarithmic non-local terms necessarily appear. For instance in 
a two-point function they appear in the combination
\be
\frac{1}{\epsilon} + \log\frac{-p^2}{\mu^2},
\ee
p being the external momentum. Note that the cut provided by the log is actually 
absolutely required by unitarity. All coefficients in the chiral 
lagrangian are nominally of ${\cal O}(N_c)$.
Loops are automatically suppressed by powers of $N_c$, because $f_\pi^2 \sim N_c$
appears in the denominator, but they are enhanced by
logs at low momenta.

We have also acquired experience from chiral lagrangians in the use of the equations of motion
in an effective theory: at any order in the chiral expansion we can use the equations of motion
derived from previous orders. For instance, using that at the lowest order
$U\Box U^\dagger - (\Box U)U^\dagger=0$ (from the ${\cal O}(p^2)$ lagrangian), one
can reduce the number of operators at ${\cal{O}}(p^4)$.

\section{Is gravity a Goldstone phenomenon?}

The 4D Einstein-Hilbert action shares several remarkable aspects with
the pion chiral lagrangian. It is a 
non-renormalizable theory as well as it is also described, considering 
the most relevant operator (we ignore here for a moment the cosmological constant), 
by a dimension two operator containing in both cases
two derivatives of the dynamical variable. Both lagrangians contain necessarily
a dimensionful constant in four dimensions: $M_P$, the Planck mass, is the counterpart of 
the cosntant $f_\pi$ in the pion lagrangian (of course the value of both constants is radically 
different). Both theories are non-linear and, finally, both describe the
interactions of massless quanta. The Einstein-Hilbert action is
\be
{\cal L}= M_P^2 \sqrt{-g} {\mathcal R} + {\cal L}_{matter},
\ee
where as just mentioned
${\mathcal R}$ contains two derivatives of the dynamical
variable which is the metric $g_{\mu\nu}$
\be
{\mathcal R}_{\mu\nu} = 
\partial_{\alpha} \Gamma^{\alpha}_{\mu \nu} 
- 
\partial_\nu \Gamma^{\alpha}_{\mu \alpha} 
+ 
\Gamma^{\alpha}_{\beta \alpha} \Gamma^{\beta}_{\mu \nu}
- 
\Gamma^{\alpha}_{\beta \nu} \Gamma^{\beta}_{\mu\alpha},
\ee
\be
\Gamma^{ \gamma }_{ \alpha \beta } = \frac{1}{2} g^{\gamma \rho} \left( \partial_{\beta }
g_{\rho \alpha } + \partial_{ \alpha } g_{\rho \beta } - \partial_{\rho} g_{ \alpha \beta } \right).
\ee
In the chiral language, the Einstein-Hilbert action would be 
${\cal O}(p^2)$ i.e. most relevant, if we omit the presence of the
cosmological constant which accompannies the identity operator. Arguably, locality,  symmetry and 
relevance in the RG sense (and not renormalizability) 
are the ones that single out Einstein-Hilbert action
in front of  e.g. ${\mathcal R}^2$.

Unlike the chiral lagrangian, the Einstein-Hilbert lagrangian, or extensions
that include higher derivative terms, has a local gauge symmetry. Indeed,
gravity can be (somewhat loosely) described as the result of promoting a global symmetry (Lorentz) 
to a local one (for a detailed discussion on the gauge structure of gravity see e.g. \cite{perc}). 
This means that the gauge symmetry that is present in gravity,
will in practice reduce the number of degrees of freedom
that are physically relevant.

Exactly like the chiral lagrangian, the  Einstein-Hilbert action requires an 
infinite number of counterterms
\be
{\cal L}= 
M_P^2 \sqrt{-g} {\mathcal R} +
\alpha_1 \sqrt{-g} {\mathcal R}^2 +
\alpha_2 \sqrt{-g} ({\mathcal R_{\mu\nu}})^2 + 
\alpha_3 \sqrt{-g} ({\mathcal R_{\mu\nu\alpha\beta}})^2 + \ldots
\ee
The divergences can be absorbed order by order by redefining the coefficients $\alpha_i$ just as 
done in the previous section for the pion effective lagrangian.
Power counting in gravity appears, at least superficially, quite similar to
the one that can be implemented in pion physics.
Of course, the natural expansion parameter is a tiny number in normal circumstances,
namely
\be
p^2/16\pi^2 M_{P}^2\quad {\rm or}\quad 
\nabla^2/16\pi^2 M_{P}^2,\qquad {\mathcal R}/16\pi^2 M_{P}^2,
\ee
making quantum effects usually quite negligible. 
There are some subtleties when matter fields are included (see \cite{don} for
a discussion).

Like in the pion chiral lagrangian non-local logarithmic pieces accompany the divergences. 
In position space they look like 
\be
\frac{1}{\epsilon} + \log\frac{\nabla^2}{\mu^2},
\ee
where $\nabla$ is the covariant derivative on symmetry grounds, $\nabla^2$ 
reducing to $-p^2$
in flat space-time. These non-localities  are due to the propagation of strictly
massless non-conformal modes, such as the graviton itself. Therefore 
they are unavoidable
in quantum gravity. Notice that the coefficient of these non local terms are entirely 
predictable from the infrared properties of gravity.

Let us use  'chiral counting' arguments to derive the relevant quantum 
corrections to Newton's law (up to a constant). The propagator at tree level
gets modified by one-loop `chiral-like' corrections
\be
\frac{1}{p^2}\quad\to\quad
\frac{1}{p^2}(1 + A\frac{p^2}{M_P^2} + B \frac{p^2}{M_P^2} \log p^2).
\ee
Consider now the interaction of a point-like particle with an static source ($p^0=0$) 
and let us Fourier transform the previous expression for the loop-corrected 
propagator in order to get the potential
in the non-relativistic limit. We recall that
\be
\int d^3 x \exp (i\vec p \vec x) \,\frac{1}{p^2} \sim \frac{1}{r},
\quad
\int d^3 x \exp (i\vec p \vec x) \,1  \sim \delta({\vec x}),\quad
\int d^3 x \exp (i\vec p \vec x) \, \log{p^2} \sim \frac{1}{r^3},
\ee
with $r=|\vec x|$. Thus quantum corrections to Newton's law are of the form
\be
\frac{GMm}{r}( 1 + K \delta(\vec x) + C \frac{G\hbar}{c^3}\frac{1}{r^2}+ \ldots ).
\ee
We have restored for a moment $\hbar$ and $c$ to make evident that $C$ is a pure number. 
The contribution proportional to $\delta(\vec x)$ 
is of course non-observable, even as a matter of principle. It comes from the 
contact divergent term (that may eventually collect contributions
from arbitrarily high frequency modes). $C$, however, is calculable. It depends only
on the infrared properties of the theory.

A long controversy regarding the value of $C$ exists in the literature\cite{khri,donbohr,qcorrs}. 
The result now accepted as the correct one, $C=41/10\pi$ \cite{bohr} is obtained 
by considering the inclusion
of quantum matter fields 
and considering the on-shell scattering matrix. Note that quantum corrections make gravity 
more attractive (by a really tiny amount) at long distances than
predicted by Newton's law.  
In addition to quantum corrections there are post-newtonian classical 
corrections that are not discussed here (see \cite{don}). 

There are in the literature definitions of an ``effective'' or ``running'' Newton constant \cite{perc2}. 
A class of diagrams is identified that dresses up $G$ and turns it into 
a distance (or energy)-dependent constant $G(r)$. Unfortunately it is not clear
that these definitions are gauge invariant; only physical observables 
(such as a scattering matrix) are guaranteed to be. Nevertheless
the renormalization-group analysis derived from this ``running'' coupling constant are of course 
very interesting and may bear relevance to the issue of asymptotic safety mentioned in
the introduction.

\section{A two-dimensional toy model}

We have given in the previous sections arguments why the Einstein-Hilbert 
action could be viewed
as the most relevant term, in the sense of the renormalization-group, 
of an effective theory describing the long distance behaviour of some
underlying dynamics. 

Here we want to pursue this line of thought further.
As a logical possibility, without making any particularly strong
claim of physical relevance, we shall investigate a formulation inspired as much as
possible in the chiral symmetry breaking of QCD. It should have the following
characteristics:

\begin{enumerate}
\item No a priori notion of metric should exist, only an affine connection
defining parallel tranport of tangent vectors $v^a$ on a manifold.
\item The lagrangian should be manifestly independent of the field $g_{\mu\nu}(x)$.
\item The graviton field should appear as the Goldstone boson of a suitably broken global symmetry.
\item The breaking should be triggered by a fermion condensate.
\end{enumerate}
A model along these lines was considered some time ago by Russo 
and others\cite{ru}. Our proposal appears to be perturbatively renormalizable 
and leads to finite calculable predictions, unlike the one in \cite{ru}.

As announced we seek inspiration in the effective lagrangians of QCD at long distances.
A successful model for QCD is the so-called chiral quark model\cite{ERT}. Consider
the matter part lagrangian of QCD with massless quarks (2 flavours)
\be
{\cal L}= i\bar \psi \not\! \partial \psi 
= i\bar \psi_L \not\! \partial \psi_L
+i\bar \psi_R \not\! \partial \psi_R.
\ee
This theory has a global $ SU(2)\times SU(2)$ symmetry that forbids
a mass term $M$. However after chiral symmetry breaking pions appear and they must be
included in the effective theory. Then it is possible to add
the following term
\be
- M \bar\psi_L U \psi_R -M \bar\psi_R U^\dagger\psi_L, 
\ee 
that is invariant under the full global symmetry $\psi_L\to L\psi_L,\ \psi_R \to R\psi_R,\
U\to LUR^\dagger$.

Chiral symmetry breaking is triggered by 
a non-zero fermion condensate $
<\bar\psi\psi>\neq 0 $.
In order to determine the value of this condensate, and in particular whether it is zero or not, 
one is to solve a `gap'-like equation in some modelization of QCD, or on the lattice.
The final step is to integrate out the fermions using
the self-generated effective mass as an infrared regulator. This reproduces the chiral 
effective lagrangian discussed in the first section, although the low-energy constants $\alpha_i$ obtained 
in this way are not necessarily the real ones, as the chiral quark model is only a 
modelization of QCD.

We shall use the Euclidean conventions. Our idea is to find out
a two-dimensional field theory with the characteristics outlined above.
There is only one possible 'kinetic' term bilinear in fermions 
that is invariant under Lorentz $\times$ {\it Diff} (actually $SO(D)$ rather than Lorentz) 
and it is local and hermitian\footnote{Actually what we really should require is that
the continuation to Minkowski space is hermitian}. It is 
\be 
i\bar{\psi}_a \gamma^a \nabla_\mu \psi^{\mu} + i\bar\psi^\mu\gamma^a\nabla_\mu \psi_a.
\label{kinetic}\ee
To define $\nabla_\mu$ we only need an affine connection 
\be
\nabla_{\mu} \psi^{\mu} = \partial_\mu \psi^{\mu} +   \omega^{a b}_{\mu}
  \sigma_{a b} \psi^{\mu} + \Gamma^{\nu}_{\mu \nu} \psi^{\mu}
\ee
Here $a,b...$ are tangent space indices, while $\mu, \nu,...$ are world indices.
The coordinates on the manifold shall be denoted by $x^\mu$ and of course there is no way of raising
or lowering indices because there is no metric. Only $\delta_{ab}$ as invariant
tensor of the tangent space is admisible. $\psi_a$ and $\psi^\mu$ are independent
spinor fields. The field $\psi^\mu$ is expected to have a spin 1/2 as well
as 3/2 component. No attempt has been made to project out the 1/2 component.  

Note that no metric is needed at all
to define the action if we assume that $\psi^\mu$ behaves
as a contravariant spinorial vector density under {\it Diff}. Then,
$\Gamma^\mu_{\nu\rho}$ does not enter in the covariant derivative, only the 
spin connection $\omega^{a b}_{\mu}$.
If we keep this spin connection fixed, i.e. we do not consider it
to be a dynamical field for the time being, there is no invariance under 
general coordinate transformations, but only under the global group
$G=SO(D)\times GL(D)$. Notice once more that the spin connection is the 
only geometrical quantity introduced.

We would like to find a non zero value for the fermion condensate
\be
<\bar \psi_a \psi^\mu + \bar \psi^\mu \psi_a> \sim A_a^\mu  \neq 0.
\ee
Because the broken theory has still the full symmetry 
it is of course irrelevant in which 
direction the condensate points; all the vacua will be equivalent. 
We can choose $A_a^\mu = \delta_a^\mu$ without loss
of generality. 

Along with the breaking a large number of Goldstone bosons are produced. 
The original symmetry group
$G=SO(D)\times GL(D)$ has $\frac{D(D-1)}{2}+D^2$ generators. After the breaking 
$G\to H$, with $H=SO(D)$ there are $D^2$ broken generators, as expected. 
It remains to be seen how many of those actually couple to physical states.

In order to trigger the appeareance of a vacuum expectation value 
we have to include some dynamics to induce the
symmetry breaking. The model we propose is to add the interaction piece
\begin{equation}
  S_I = \int d^4 x (iB_\mu^a( \bar{\psi}_a \psi^{\mu} + \bar{\psi}^{\mu} \psi_a)
   + c \det (B^a_{\mu}))\label{int}
\end{equation}
Note that the interaction term also behaves as a density thanks to the
covariant Levi-Civita symbol hidden in the determinant of $B^a_\mu$ so no
metric is needed. Note that (\ref{int}) is non-hermitian, but the continuation
to Minkowski is: $B_\mu^a$ upon continuation changes like an Euclidean mass
does $B_\mu^a \to i B_\mu^a$. Since the field $B_\mu^a$ is auxiliary, it is clear
that we are dealing with a four-fermion interaction; fermions are the only dynamical
fields.

\subsection{Equations of motion}

If we consider the equation of motion for the auxiliary field $B^a_\mu$ 
we get
\be
<\bar \psi_a \psi^\mu + h.c. > = -i c \epsilon^{\mu\nu}\epsilon_{ab} B^b_\nu.
\label{eom1}
\ee
We conjecture the field $B_\mu^a$ to correspond to the
zwei-bein, $e^a_\mu$, up to a (dimensionful) constant.

Making use of this equation of motion, the interaction term 
in 2D corresponds to a four-fermion interaction\footnote{Although 
this would take us too far away, note
that this is reminiscent of a instanton-generated interaction. We are
grateful to C. G\'omez for a discussion on this subject}.
\be
\epsilon_{\mu\nu}\epsilon^{ab}(\bar \psi_a \psi^\mu + \bar \psi^\mu \psi_a)
(\bar \psi_b \psi^\nu + \bar \psi^\nu \psi_b)
\ee
This can be integrated over the manifold without having to 
appeal to a measure if we assume that $\psi^\mu$ is a spinorial
density. Note that if $\langle \bar \psi_a \psi^\mu + 
\bar \psi^\mu \psi_a\rangle$ acquires a v.e.v. it is possible to write
new operators. 

The equations of motion for the fermion fields are:
\be
\gamma^a \nabla_\mu \psi^\mu + B_\mu^a \psi^\mu=0
\ee
and
\be
\gamma^a\nabla_\mu \psi_a + B_\mu^a \psi_a=0.
\ee
Note that after use of the equations of motion the lagrangian itself
reduces to the term $c \det(B_\mu^a)$.

\subsection{Energy-momentum tensor and symmetries}

Altough the above theory is 'topological' inasmuch as it is described
by an action that does not contain a metric (albeit it depends on a
connection), the energy-momentum tensor
understood as the Noether currents of translation invariance is non-vanishing
\be
T^\mu_\nu= i \bar\psi^\mu\gamma^a\partial_\nu\psi_a 
+i \bar\psi_a\gamma^a\partial_\nu\psi^\mu -\delta^\mu_\nu L.
\ee
Note that no metric is needed to define $T^\mu_\nu$. In the absence of 
the external connection $T^\mu_\nu$ is traceless as expected given that 
the theory is 
formally conformal, but we will see later that it will not remain so 
at the quantum level as anomalous dimensions develop.

The free action (\ref{kinetic}), without considering the 
interaction term, is invariant under the symmetry 
\be
\psi_a \to \psi^\prime_a= (\delta_a^b - \frac1D \gamma_a \gamma^b)\psi_b.
\ee
Another invariance of the free action is provided by redefining, in Fourier
space,
\be
\psi^\mu(k) \to \psi^{\prime \mu}=P^\mu_{\nu}\psi^\nu(k),
\ee
where $k_\mu P^\mu_{\nu}=0$. These two invariances difficult considerably the
heat kernel derivation of an effective action for the field $B_\mu^a$ that
will be discussed below.

\subsection{Free propagator and renormalizability}

Note the peculiar 'free' kinetic term $\gamma^a \otimes k_\mu$. It is 
of course reminiscent of Dirac equation, but it is not quite identical
(Dirac needs a metric or a  n-bein to be defined).
In the next subsection we will see that after the introduction of the
interaction term $~\sim$ det$B$, the field $B_\mu^a$ will indeed develop 
a v.e.v. that we conventionally take to be
\be
\langle B_\mu^a \rangle = M \delta_\mu^a.
\ee
Any other direction would be equivalent. The only substantial fact is whether
$M$ is zero or not.
Via (\ref{eom1}) this v.e.v for $B_\mu^a$ translates into a v.e.v for 
$\bar \psi_a \psi^\mu + \bar \psi^\mu \psi_a$
Fom (\ref{int}) we see that the scale $M$ plays the role of a
dynamically generated mass for the fermions (not unlikely the 'constituent
mass' in chiral dynamics, except that here it will be possible, as we will
see, to determine exactly its relation to the fundamental parameters of
the model).

Below we write explicitly in two dimensions the bilinear operator
acting on the fermion fields. Considered as a matrix, we shall not distinguish at this point
between tangent and world indices (we then use indices in the
middle of the alphabet: $i, j, k,...$).
\begin{equation}
     {\Delta}_{ij} = \left( \begin{array}{cccc}
      i B_{11} &  k_1 &i B_{12} &  \hspace{0.25em} k_2\\
        \hspace{0.25em} k_1 &i B_{11} &  \hspace{0.25em} k_2 & iB_{12}\\
      i B_{21} & - i k_1 & iB_{22} & - i k_2\\
       i k_1 &iB_{21} & i k_2 &i B_{22}
     \end{array} \right).
\end{equation}
When $B_{ij}$ develops a v.e.v., $B_{ij}= M\delta_{ij}$, this reads
\begin{equation}
     {\Delta(k)}_{ij} = \left( \begin{array}{cccc}
       iM &  k_1 & 0 &  \hspace{0.25em} k_2\\
        \hspace{0.25em} k_1 & iM &  \hspace{0.25em} k_2 & 0 \\
       0 & - i k_1 & iM & - i k_2\\
       i k_1 & 0 & i k_2 & iM
     \end{array} \right). \label{kinope}
\end{equation}
The inverse of this matrix will give the propagator of the fermion field.
It can be written (in any number of dimensions) as
\be
\Delta^{-1}(k)_{ij}=\frac{-i}{M}\left(\delta_{ij}-\frac{\gamma_i(\not\! k -i M )k_j}
{k^2 + M^2}\right),
\ee
with $k^2= \sum_i k_i^2$. The covariance of the results, not
evident at all from these expressions, will be discussed
in the next section.

This is an appropriate point to discuss the renormalizability of the model.
Naively, because the coupling constant $c$ is dimensionless in 2D, we would
expect the model to be renormalizable. However this expectation is jeopardized 
by the behaviour of the propagator. Indeed the diagonalization of
(\ref{kinope}) gives as eigenvaules: $M$ (twice), $k+iM$ and $k-iM$. Therefore
the propagator does not behave, in general, as $1/k$ and therefore the usual
counting rules simply do not apply. 

There is however a further twist to the issue
of renormalizability. The model proposed does not contain a metric and therefore
the number of counterterms that one can write is extremely limited. For instance, 
a mass term for the $B$ field is impossible. Higher dimensional operators would
require powers of $\sqrt{g}$ to preserve the $Diff$ invariance that the model
has (when $w$ is a dynamical variable), but there is no metric. In fact the metric
will be generated {\em after} the breaking, but the counterterms of a field theory
do not depend on whether there is spontaneous breaking of a global symmetry or not. 
In summary, the
lack of counterterms make us believe that the theory is renormalizable after
all. Indeed this expectation is supported by an explicit one-loop calculation 
(see section 6) where
the only divergence that appears can be absorbed by a redefinition of $c$. We find
this quite remarkable.

\subsection{Gap equation}

If $w_\mu=0$ then one can use homogeneity
and isotropy arguments to look for constant solutions of the gap equation
associated to the following effective potential
\be 
V_{eff} = c\, \det (B^a_{\mu}) - 2 \int \frac{d^D k}{(2 \pi)^D}
   {\rm tr\;} (\log ( \gamma^a_{} k_{\mu} +i B^a_{\mu})).
\ee
The factor 2 is due to the fact that $\psi^\mu$ and $\psi_a$ are independent
degrees of freedom. By deriving w.r.t $B_\mu^a$ the 
extrema of $V_{eff}$ are found from the equation
\be c~ n ~\epsilon_{a a_2 \ldots .a_n} \epsilon^{\mu \mu_2 \ldots . \mu_n}
   B^{a_2}_{\mu_2} \ldots .B^{a_n}_{\mu_n} - 2~ i~ {\rm tr\;} \int \frac{d^D k}{(2
   \pi)^D} (\gamma\otimes k +i B)^{- 1}\vert^\mu_a = 0.
\ee
In 2D this equation is particularly simple
\be
c \epsilon_{ab}\epsilon^{\mu\nu}B^b_\nu - i~{\rm tr\;} \int \frac{d^D k}{(2
   \pi)^D} ( \gamma\otimes k + iB)^{- 1}\vert^\mu_a = 0.
\ee
The `gap equation' to solve for constant values of $B_{i j}$ is
\be
c B_{i j} + \frac{1}{2\pi}B_{i j}\log \frac{\det B}{\mu^2}= 0.
\ee
A logarithmic divergence has been absorbed in $c$. 
Notice that the equations are invariant under the permutation
\be B_{i j} \rightarrow B_{\sigma (i) \sigma (j)}, k_i \rightarrow k_{\sigma
   (i)}, \quad \sigma \epsilon S_2.
\ee
This equation has 
a non-trivial solution  that we can always choose, as indicated before,
to be $B_{ij} \sim \delta_{ij}$. We thus see that the dynamical mass
for the fermions is indeed generated hence justifying {\em a posteriori} the propagator
introduced in the previous subsection. The solution for the dynamical mass is
\be
M=\mu e^{-\pi c(\mu)}.\label{ggap}
\ee
Plugging this back in the effective potential we obtain 
\be
V_{eff}= - \frac{\mu^2 e^{-2\pi c(\mu)}}{2\pi}.
\ee
Upon continuation to  Minkowski space-time this term is to be 
identified with the cosmological constant. Because
when rotating to Minkowski space $V\to -V$, the cosmological constant is
positive. At this level $M$ 
is an observable and as such it should be a
renormalization-group invariant. This is guaranteed if $c$ runs according to the rather
trivial beta function
\be
\mu \frac{dc}{d\mu}= \frac{1}{\pi}.
\ee
Note that the coefficient of this term is related to the coefficient of the logarithmic
divergence and hence it is universal.

The above effective potential and ensuing gap equation are exact in the 
limit where the number of fermions, $N$,  is infinite. In fact we expect that it
is exact {\em only} in this limit, as in 2D the phenomenon of spontaneous
breaking of a continuous symmetry can take place only in the $N=\infty$ limit. 

For non-zero connection ($w_\mu\neq 0$) the gap equation is not applicable and
one needs to derive the full effective action. Then one would minimize
the fiels $B_\mu^a$ as a function of $w_\mu$. This is discussed in the next section.

\section{Derivation of the effective action}

Let us now attempt to derive the effective action for the fields $B_\mu^a $ and
the external affine connection  $w_\mu$ that eventually we will allow to
become a dynamical variable too. Hereafter we want to
perform a double minimization with respect to these fields. 
This will be an exact procedure for $N=\infty$ and provide a guidance 
in the general case. Of course the really interesting question 
is what happens for $D>2$.

We would expect that this double minimization will provide us with two equations
whose meaning  would be schematically the following: One of them 
would provide a relation between
the field $B_\mu^a$ ( associated to the zwei-bein) and the affine 
connection $w_\mu$. If the present model is to describe in its broken phase
2D gravity, this relation  would be analogous
to the relation of compatibility between the metric and the connection 
that appears when the Palatini formalism\cite{palatini} is used in General Relativity and the
equations of motion for the connection $w_\mu$ are derived.
The remaining equation, after imposition of the previous compatibility condition,  should 
then be equivalent to Einstein's equations.

However, in 2D gravity is rather peculiar and indeed the condition
\be
w_\mu^{ab}= e_\nu^a\partial_\mu E^{\nu b} + e_\nu^a E^{\sigma b}\Gamma^\nu_{\sigma\mu},
\ee
where $E_a^\mu$
is the inverse zwei-bein $E_a^\mu e^b_\mu=\delta^b_a$,
holding in any number of dimensions, does not follow in 2D from
any variational principle (see e.g. \cite{xin}). There are several ways to 
understand this fact, but perhaps the 
simplest one is to realize that Einstein-Hilbert in 2D depends on
$w_\mu$ only through the two-form {\bf dw} which is linear in the affine
connection $w_\mu$. In fact the scalar curvature term $\sqrt{g}{\cal R}$ does not contain
in 2D any coupling between $g_{\mu\nu}$ and $w_\mu$. Adding higher derivatives 
does not really help as the Riemann tensor contains only an independent component
that can be ultimately related to the scalar curvature. 
We shall see below that this peculiarity of two-dimensional gravity is 
faithfully reproduced in our proposal.

The starting point of the derivation of the effective action 
is the differential operator
\be
{D}_{\mu}^{a}= \gamma^a (\partial_\mu +  w_\mu\sigma_3) + B_\mu^a.
\ee
We consider the expansion around a fixed background preserving $SO(D)$ but
not the full symmetry group $G$. We will take $B_\mu^a= M \delta_\mu^a$ where
$M$ will be determined via the gap equation discussed
in the previous section, which corresponds to a solution of the equation of motion
at the lowest order in a weak field and derivative expansion, in the spirit
of effective lagrangians. To go beyond this approximation we have to consider 
$x$-dependent fluctuations around this vacuum and include the external field $w_\mu$.
We shall decompose
\be
B_\mu^a = \xi_{L~b}^{a}\bar{B}_{\nu}^{b}\xi_{R~\mu}^{-1\nu},
\ee
where $\xi_{L}\in SO(D)$, $\xi_{R}\in GL(D)$ and $\bar B_\mu^a$ is a solution
of the gap equation, $M\delta_\mu^a$ in our case. 
It is technically advantageous to absorb the matrices $\xi_L$ and $\xi_R$
in the fermion fields (in QCD this is the so-called 'constituent' quark basis
\cite{ERT}). Then the differential operator to deal with will be
\be
\mathcal{D}_{\mu}^{b}=\xi_{La}^{\dagger~ b}\gamma^{a}(\partial_{\rho}+
w_{\rho}\sigma_{3})\xi_{R~\mu}^{\rho}+\bar{B}_{\mu}^{b}.
\ee

To evaluate the effective action generated by the integration of the fermion fields 
one possibility is to
write the log of the fermion determinant as
\begin{equation}
W=-\frac{1}{2}\int_{0}^{\infty}\frac{dt}{t}\text{tr}
\left<x|e^{-t X}|x\right>, \label{heatkern}
\end{equation} 
where
\be
X_{\mu\nu} \equiv {\cal M}^\dagger {\cal M},
\ee
with
\begin{equation}
{\cal M}=  \mathcal{D}_{\nu}^{b},\qquad {\cal M}^{\dagger}= -\mathcal{D}_{\mu b}
\end{equation}
and
\begin{equation}
\mathcal{D}_{\mu}^{b}=\xi_{La}^{\dagger~ b}\gamma^{a}(\partial_{\rho}+w_{\rho}\sigma_{3})\xi_{R~\mu}^{\rho}
+\bar{B}_{\mu}^{b},\,\,
\mathcal{D}_{\nu b}=\xi_{R\nu}^{\dagger~ \sigma}(\partial_{\sigma}-w_{\sigma}\sigma_{3})\gamma_{a}\xi_{L~b}^{a}
-\bar{B}_{\nu b}.
\end{equation}
$X_{\mu\nu}$ has both world and Dirac indices (the latter not explicitly written).
Note that as previously discussed ${\cal M}$ is not hermitean, but of course
$X_{\mu\nu}= {\cal M}^\dagger {\cal M}$ is. We could have also considered the determinant of ${\cal M}{\cal M}^\dagger$ which is
of course identical, but it is important to mantain a covariant appeareance as long as
possible (note that there is no metric so far and no way of lowering or raising
indices). The final result has to be of course covariant, since our starting point is, but
using, as we shall do, a plane basis to evaluate the traces in the heat kernel expansion
breaks in principle this covariance in intermediate steps.

Once $W(w,B)$ is known we can
differentiate with respect $w_\mu$ and obtain the relation between the zwei-bein and the
spin connection using the logic behind the Palatini formalism.
 
The starting point of the heat kernel derivation is the evaluation of
\begin{equation}
\text{tr}\langle x\vert e^{-tX_{\mu\nu}}\vert x\rangle=
\frac{1}{t^{D/2}}\int\frac{d^Dk}{(2\pi)^D}
\text{tr}~e^{[-D\xi_{R\mu}^{\top~\sigma}\xi_{R\nu}^{\rho}k_{\sigma}k_{\rho}
+i\sqrt{t}\mathcal{D}_{\mu b}\xi_{L~~a}^{-1b}\gamma^{a}k_{\rho}\xi_{R\nu}^{\rho}
+i\sqrt{t}\xi_{R\mu}^{\top~ \sigma}k_{\sigma}\gamma_{a}\xi_{L~~b}^{-1a}\mathcal{D}_{~\nu}^{b}
-t X_{\mu\nu})]}\\
\end{equation}
where for convenience we have rescaled $k_{\mu}$ and a plane wave basis 
resolution of the identity has been used. For simplicity
let us call the exponent on the r.h.s. of the previous equation $X(\sqrt{t})$. Then
the way to proceed is to expand 
the exponential $e^{X(\sqrt{t})}$ in powers of $\sqrt{t}$. Only even powers of $\sqrt{t}$ (and thus of 
$k$) will contribute at the end to the series, so the first non-trivial term will be of order 
$t$. We define
\begin{equation}
F_{n}\left(X(0),\dot{X}(0),\ddot{X}(0)\right)\equiv \frac{d^{(n)}}{(d\sqrt{t})^n}e^{X(\sqrt{t})}\mid_{\sqrt{t}=0}
\end{equation}
then
\begin{equation}
\begin{aligned}
\text{tr}\left<x|e^{-t X_{\mu\nu}}|x\right>\propto ~ &\text{tr}~\sum_{n}F_{n}\frac{(\sqrt{t})^{n}}{n!}
=F_{0}+\frac{t}{2}F_{2}+\frac{t^2}{24}F_{4}+\mathcal{O}(t^4)
\end{aligned}
\end{equation}
This expansion is quite tedious and to perform it we used repeatedly the well known formula
\be
\frac{d}{dt}e^{A(t)}=\int_0^1 da~e^{(1-a)A(t)} \frac{dA(t)}{dt} e^{a A(t)}.
\ee
Note that the invariances discussed in the previous section introduce
zero modes in the exponent and hence integrals that are not damped for large
values of the momentum $k$. Of course they are no true zero modes
of the full theory, just of the kinetic term, but the technical complications
that they bring about are notable. 

However, it is pleasant to see that a formally covariant result
emerges. If we neglect $w_\mu$ and we take the matrices $\xi$ to
be constant it is not difficult to see that the lowest non-trivial
order of the heat kernel calculation gives
\begin{equation}
W=\frac{3M^2}{16\pi} 
\int d^{2}x\sqrt{\text{Det}[(\xi_{R\,\mu}^{\sigma}\xi_{R\,\mu}^{\dagger\rho})^{-1}]},
\end{equation}
where a summation over $\mu$ is to be understood and where
$M^2$ is the dynamically generated mass.
This is just a cosmological term with 
$g^{\sigma\rho}= \sum_\mu \xi_{R\,\mu}^{\sigma}\xi_{R\,\mu}^{\dagger\rho}$.
One can likewise verify that other pieces in the effective action
are covariant. The coefficient of the cosmological constant term obtained
at the lowest order in the heat kernel expansion does not
agree with the one obtained through the gap equation. We shall see later why this is so.

Since the most general metric in two dimensions is conformally flat
we can reconstruct the full covariant action from this particular
choice. This simplifies notably the derivation of the effective action. 
We take $B_{~\mu}^{a}(x)=\xi_{Lb}^{a}\bar{B}_{~\rho}^{b}\xi_{R~~
\mu}^{-1\rho}(x)=M\phi^{-1}\delta_{\mu}^{a}$. 
The expressions that follow are specific to this gauge.

At second order in the heat kernel expansion (order $(\sqrt{t})^2$) the corresponding 
piece of the effective action reads
\begin{equation}
\begin{aligned}
W^{(2)}=\int d^{2}x~\phi^{-2}
&\left[\frac{3M^2}{16\pi} \left(
 \frac{2}{\epsilon}-\gamma-\text{log}\left(\frac{M^2}{\mu^2}\right)
 +\text{log}(8\pi)+4\right)\right.\\
&\left. +\frac{(\partial_{\mu}\phi)^2}{4\pi}\left(\frac{2}{\epsilon}
-\gamma -\text{log}\left(\frac{M^2}{\mu^2}\right)
 +\text{log}(8\pi)- \frac{5}{3}\right)\right.\\
 &\left. +\frac{w^2\phi^2}{4\pi}\left(\frac{2}{\epsilon}
-\gamma -\text{log}\left(\frac{M^2}{\mu^2}\right)
 +\text{log}(8\pi)\right)\right],
\end{aligned}
\end{equation}
where $D=2 - \epsilon$.
We can take one step further and calculate the contribution to order $t^2$ 
\begin{equation}
\begin{aligned}
W^{(4)}=\int d^{2}x~\phi^{-2} &\left[-\frac{3M^2}{32\pi}\left(\frac{2}{\epsilon}+\text{log}\left(8\pi\right)-
\text{log}\left(\frac{M^2}{\mu^2}\right)
-\gamma +\frac{5}{18}\right)\right.\\
&\left. -\frac{\phi^4(\partial_{\mu}w_{\mu})^2}{4\pi M^2}+\frac{\phi^3 \left(2w_{\nu}\partial_{\mu}\phi\partial_{\mu}w_{\nu}-3w_{\mu}\partial_{\mu}\phi\partial_{\nu}w_{\nu}\right)}{3\pi M^2}\right.\\
&\left. \frac{\phi^2\left(\partial_{\mu}\phi\right)^2w^2}{6\pi M^2}-\frac{w_{\mu}w_{\nu}\partial_{\mu}\phi\partial_{\nu}\phi}{\pi M^2}\right.\\
&\left. +\frac{\phi^2 w^2}{8\pi}-\frac{\phi^4 w^4}{4\pi M^2} -\frac{5(\partial_{\mu}\phi\partial_{\mu}\phi)}{48\pi}\right.\\
&\left. +\frac{\phi}{M^2}\frac{\partial_{\mu}\phi\partial_{\nu}\phi\partial_{\mu}\partial_{\nu}\phi}{ 3\pi}+
\frac{\phi}{M^2}\frac{\partial_{\mu}\phi\partial_{\mu}\phi\partial_{\nu}\partial_{\nu}\phi}{ 15\pi}\right.\\
&\left. -\frac{7\phi^3}{M^2}\frac{\partial_{\mu}\partial_{\mu}\partial_{\nu}\partial_{\nu}\phi}{ 60\pi}-\frac{\partial_{\mu}\phi\partial_{\mu}\phi\partial_{\nu}\phi\partial_{\nu}\phi}{ 5\pi M^2}
\right]
\end{aligned}
\end{equation}
The calculation of the fourth-order coefficients in the heat kernel expansion just shown
is already a formidable task and we will not attempt to go beyond. 

If we look at the results of the expansion at second order it is interesting to see 
that the terms that are generated are the ones
expected from the point of view of general relativity. 
There is a cosmological term (proportional to $\phi^{-2}$, which in covariant
form corresponds to $\sqrt{g}$), and 
a Liouville term (proportional to $\phi^{-2} (\partial_\mu\phi)^2$, which in
covariant form is non-local: $\sqrt{g}{\cal R}\ \nabla^{-2}\sqrt{g}{\cal R}$). In addition there
is a term proportional to $w^2$ (which once written in a covariant form would be
$\sqrt{g}g^{\mu\nu}w_\mu w_\nu$).
Note that the Einstein term itself is topological in 2d and it is not expected to show up.
However, in spite of these satisfactory results, we notice that 
the cosmological term does not quite coincide with the one previously derived, 
via the gap equation, and the
Liouville term is apparently divergent casting doubts on the renormalizability of the model. 
We note that like in the chiral lagrangian, the effective theory still posesses the full
symmetry grup $G$.

Yet it is easy to see that the above results are by necessity incomplete. For instance, the 
same operator $\phi^{-2}$ gets a contribution from the terms of order $t$ and from $t^2$, ditto
for Liouville. This comes from the fact that because the operator $X(\sqrt{t})$ contains
terms linear in $M$ and the heat kernel expansion is effectively an expansion in inverse powers
of $M$, a given order in $t$ does not correspond to a given order in derivatives or external fields.
Therefore although the heat kernel calculation gives an interesting guidance to the form
of the effective action and it shows the reappeareance of covariance, the precise values 
of the coefficients of the different operators
cannot be extracted from it. To solve this difficulty we turn to a diagramatic calculation.

\section{Diagramatic calculation}

Let us recapitulate. The heat-kernel calculation is plaged by two problems. The first one is related 
to the zero modes of the kinetic term, which increase considerably the difficulty of the calculations. 
The other one lies in the fact that the expansion is ill-defined in the sense of relevance of the subsequent orders. 
In a way, the heat-kernel fails to provide exact coefficients for the different operators 
but gives an accurate catalogue of the possible terms one could expect.

In this section we derive the Feynman rules of our toy model and proceed to calculate 
the exact contributions of the zero, one and two-point functions. As will be shown, we obtain 
finite contributions except for the cosmological term which nevertheless can be renormalized. 
The theory appears to be perfectly renormalizable in spite of the apparent bad power counting (due 
to the zero modes of the propagator).

\subsection{Feynman rules}

We start by writing the generating functional of the theory in the conformal gauge in Euclidean space 
from which we can read off the Feynman rules for the one and two point functions we are interested in.
We know that the diagrammatic expansion is not covariant, but once we have convinced ourselves
that covariance is recovered, we can use this method for identifying specific coefficients. In this section
it will be convenient to express the conformal gauge in the form 
\be
B_\mu^a(x) = M e^{-\sigma(x)/ 2} \delta_\mu^a.
\ee
The first term in the expansion of the exponential provides the dynamically generated mass for the fermions.
Incidentally, this formalism is clearly quite reminiscent of chiral dynamics.

The interaction vertices are\\

\begin{equation}\label{vert}
\begin{aligned}
\parbox{30mm}{
\includegraphics[]{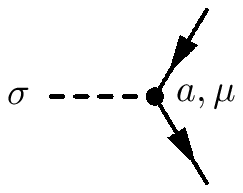}}
&\qquad i\frac{1}{2}M\delta_{\mu}^{a}\\
\\
\parbox{30mm}{
\includegraphics[]{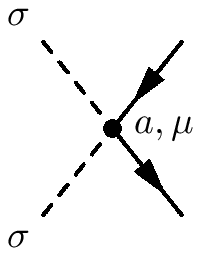}}
&\qquad -i\frac{1}{8}M\delta_{\mu}^{a}\\
\\
\parbox{30mm}{
\includegraphics[]{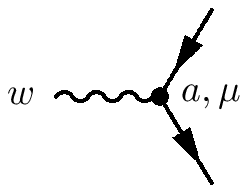}}
&\qquad -i\gamma^{a}\sigma_{3}
\end{aligned}
\end{equation}

\subsection{Zero, one and two point functions}

With the rules described in the previous section and the propagator derived in section (4.3) 
we can calculate the exact contributions of the zero, one and two point functions of the theory.
Since the theory is non-standard, and it has a non-familiar set of Feynman rules, we 
will provide below the diagrams, after transcribing the Feynman rules, and the final result.
Note that because there are two species of fermions the result from the Feynman diagrams has to be
multiplied by a factor 2. 
Let us first consider 1PI diagrams containing the $\sigma$ field as external one
\begin{equation}\label{vac}
\begin{aligned}
\parbox{15mm}{
\includegraphics[]{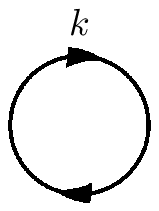}}
=&- \text{Tr}\left[\int \frac{d^{D}k}{(2\pi)^D}\Delta^{-1}(k)_{~a}^{\mu}\delta_{~\mu}^{a}\right]\\
=&\frac{iM}{2\pi}\left(\frac{2}{\epsilon}-\gamma-\text{log}\left(\frac{M^2}{\mu^2}\right)+\text{log}(4\pi)\right)\\
\\
\parbox{27mm}{
\includegraphics[]{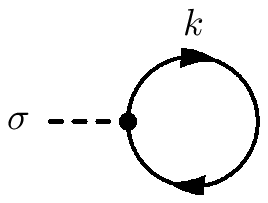}}
=&-\text{Tr}\left[\int \frac{d^{D}k}{(2\pi)^D}\Delta^{-1}(k)_{~a}^{\mu}\frac{i}{2}M\delta_{~\mu}^{a}\right] \\
=&-\frac{M^2}{2\pi} \frac12 \left(\frac{2}{\epsilon}-\gamma-\text{log}\left(\frac{M^2}{\mu^2}\right)
+\text{log}(4\pi)\right)\\
\\
\parbox{24mm}{
\includegraphics[]{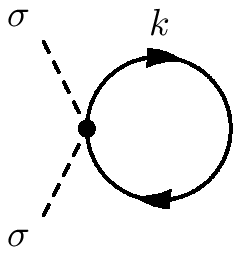}}
=&-\text{Tr}\left[\int \frac{d^{D}k}{(2\pi)^D}\Delta^{-1}(k)_{~a}^{\mu}\frac{-i}{8}M\delta_{~\mu}^{a}\right]\\
=&\frac{M^2}{2\pi}\frac18\left(\frac{2}{\epsilon}-\gamma-\text{log}\left(\frac{M^2}{\mu^2}\right)
+\text{log}(4\pi)\right)
\end{aligned}
\end{equation}
There is another diagram with two external scalar legs
\begin{equation}\label{liou}
\begin{aligned}
\parbox{40mm}{
\includegraphics[]{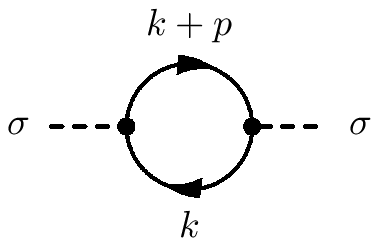}}
=&-\text{Tr}\left[\int \frac{d^{D}k}{(2\pi)^D}\frac{i}{2}M\delta_{~\mu}^{a}\Delta^{-1}(k)_{~b}^{\mu}\frac{i}{2}M\delta_{~\nu}^{b}\Delta^{-1}(k+p)_{~a}^{\nu}\right]\\
=&\frac{M^2}{4\pi}\frac14\left[\frac{2}{\epsilon}-\gamma-\text{log}\left(\frac{M^2}{\mu^2}\right)+\text{log}(4\pi)-2\right]
-\frac{1}{48\pi} p^2\\
\end{aligned}
\end{equation}
from the $M^2$-terms in (\ref{vac}) and (\ref{liou}) we can already infer 
the total contribution to the cosmological term
\begin{equation}\label{cosmo}
\frac{M^2 e^{-\sigma}}{4\pi}\left(\frac{2}{\epsilon}-\gamma-\text{log}\left(\frac{M^2 e^{-\sigma}}{\mu^2}\right)
+\text{log}(4\pi)+1\right)
\end{equation}
The divergence can be absorbed in the redefinition of the coupling constant, $c$. This
result fully agrees with the one derived via the gap equation previously. In addition we observe that 
the $p^2$ piece in the last diagram will correspond in position space to the Liouville term. As it can be seen
it is finite.

Next we look at the two point function that mixes a $\sigma$-field with $w$-field. 
This could yield a ${\cal R}$-type term but since in two dimensions gravity is topological 
we do not expect to see such term. Indeed, the diagram gives zero
\begin{equation}\label{raro}
\begin{aligned}
\parbox{40mm}{
\includegraphics[]{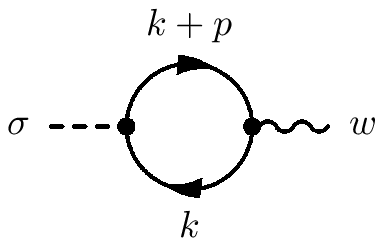}}
=&-\text{Tr}\left[\int \frac{d^{D}k}{(2\pi)^D}\frac{i}{2}M\delta_{~\mu}^{a}\Delta^{-1}(k)_{~b}^{\mu}(-i\gamma^{b}\sigma_{3})\Delta^{-1}(k+p)_{~a}^{\nu}\right]\\
=& \ 0
\end{aligned}
\end{equation}
Finally we calculate the last of the two point functions possible. Again we obtain a finite result
\begin{equation}\label{R2}
\begin{aligned}
\parbox{40mm}{
\includegraphics[]{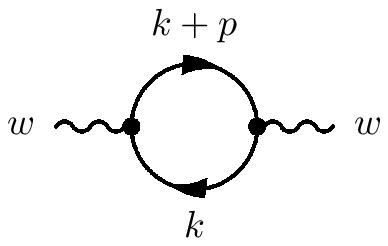}}
=&-\text{Tr}\left[\int \frac{d^{D}k}{(2\pi)^D}(-i\gamma^{a}\sigma_{3})\Delta^{-1}(k)_{~b}^{\mu}(-i\gamma^{b}\sigma_{3})\Delta^{-1}(k+p)_{~a}^{\nu}\right]\\
=& \ \frac{\delta^{\mu\nu}}{2\pi} -\frac{p^\mu p^\nu}{6\pi M^2}  
\end{aligned}
\end{equation}
We see with relief that even if the ultraviolet behaviour of each and one of the integrals is very bad, the final
result hints to the renormalizability of the theory. After renormalizing 
the only coupling constant $c$ of the theory the final result is perfectly finite.

\subsection{Effective action}
Let us now put all the pieces together and use  the lowest order equations of motion for the
field $B_\mu^a$; or what is tantamount, for the dynamically generated mass $M$, to write the
effective action. The result is 
\begin{equation}\label{efac}
S_{\text{eff}}=\int d^2x~\left[-\frac{M^2}{2\pi}e^{-\sigma}
+\frac{1}{24\pi}\partial_{\mu}\sigma\partial_{\mu}\sigma+
\frac{(\partial_\mu w_\mu)^2 }{3\pi M^2}-\frac{w^2}{\pi } + ...\right],
\end{equation}
with $M$ given by (\ref{ggap}).
This is our final result.

Several comments are in order. First we recall that the effective action is written
in the conformal gauge for the metric, but it is trivial to recover a full covariant form. 
Secondly, we note that there is no coupling between metric and connection, as befits the
Palatini formalism in two dimensions where, exceptionally, metric and connection are unrelated.
One can apply a variational principle to the affine connection $w_\mu$ in the above effective action, 
obtaining some equations of motion at ${\cal O}(p^2)$, but in 2D they do not provide any information
on the conformal factor $\sigma$.

One is then left with a cosmological and a Liouville term, as 
corresponds to two-dimensional gravity\cite{poly}.
The dots correspond to higher curvatures that we have not attempted to compute. In general
they will be non-zero. Notice that the expansion is valid as long as the characteristic momenta
fulfill $k > M$. Since $M$ is the mass scale related to the two-dimensional cosmological
constant, this would correspond to scales larger than the horizon.

\section{Four dimensions}

It is  almost compulsory to discuss the possible extension of these results to 
four dimensions.

There is apparently a fundamental problem in considering four dimensional theories
where the graviton is generated dynamically. If we refer to the original paper by Weinberg 
and Witten\cite{ww}, the apparent pathology of these theories lies in the fact that the energy-momentum
tensor has to be identically zero if particles with spin higher than one appear
and we insist in the energy momentum tensor being Lorentz covariant. This
result does not hold in two dimensions as it relies on angular momentum considerations that
do not apply. Of course there is no true spontaneous breaking of continuous symmetries in
two dimensions either, but this is circumvented by appealing to the large $N$ limit.

However, very preliminary results indicate that one gets from the simple toy model proposed here
a result that looks close to what one expects in four-dimensional gravity, so it is
legitimate to ask whether Weinberg and Witten result applies. We note something peculiar
in our model, namely the energy-momentum tensor derived in section 4 does not have tangent (Lorentz)
indices. In fact Lorentz indices are of an internal nature in the present approach, the connection
between Lorentz and space time indices appears only after a $n$-bein is dynamically
generated. But then one is exactly in the same situation as General Relativity where the
applicability of \cite{ww} is excluded. Thus it is not clear to us
to what extent the conditions assumed by Weinberg and Witten apply to the present model.

Of course in four dimensions we would possibly generate a graviton field, hopefully with the
same couplings dictated by General Relativity, but with all the degrees of freedom thrown in.
To see that only two of them are physical one has to go through the usual procedure of gauge
fixing.

It is worth noticing that the loop integrals of the present model 
show an even worse ultraviolet behaviour in four
dimensions. However, exactly as in two dimensions, the absence of a metric in the fundamental
theory seriously limits the number of possible counterterms. 

Then while the previous two-dimensional example is all too trivial it 
shows perfectly the general ideas. It seems conceivable to 
entertain the idea that a mechanism analogous to chiral
symmetry breaking may trigger the dynamical appeareance of some degrees of freedom that,
at the very least,  formally reproduce the Einstein-Hilbert action.

In four dimensions there are two independent dimensionful parameters in gravity: the scale
associated to the cosmological cosntant $\Lambda^{1/4}$ and the Planck mass $M_P$, which is absent in
2D. In a simple model (such as the one proposed here) it is to be expected that the two scales
are related (this may not necessarily be so if the 4D model requires additional subtractions). 
This is of course the old problem of fine-tuning associated to the cosmological constant
re-emerging; the novel aspect here being that in the present microscopic model of gravity this can be made
quantitative. Solving this fine-tuning problem was not however our primary motivation, but rather
trying to circumvent the difficulties of the Einstein-Hilbert lagrangian at the quantum level by
considering it as an effective theory and proposing a microscopic toy model.

\section{Summary}

In this work we have proposed a model where two-dimensional gravity emerges from a theory
without any predefined metric. The minimal input is provided by assuming a differential 
manifold structure endowed with
an affine connection. 

We have made an allusion in the title of this article to the
emergence of geometry and this is really what happens in the model proposed. Gravity and
distance are induced rather than fundamental concepts. At sufficiently short scales, when the
effective action does not make sense anymore, the physical degrees of freedom are fermionic.
Below that scale there is not even the notion of a smaller scale: in a sense that is the shortest
scale that can exist. 

A very important aspect of the model is that it appears to be renormalizable. All divergences
can be absorbed in the redefinition of a unique coupling constant. With the appropiate running,
dictated by the corresponding beta function, the
cosmological constant becomes a renormalization group invariant. Everything else is finite.
Of course at long distances the conformal mode is the relevant degre of freedom and this induces
new divergences which are, nevertheless, the ones characteristic of two-dimensional gravity. They
have been discussed in section 3. The renormalizability aspect of this model (and its relative
technical simplicity) is its main virtue when compared with previous proposals \cite{ru}, where
even semiquantitative discussions appear impossible. Here one is able to derive in all detail
the effective action. The derivation remains valid for supra-horizon momenta, where the theory
is governed by the induced metric modes, while at sub-horizon 
energy scales everything seems to indicate that the fundamental fermionic degrees of freedom are 
the relevant ones.

The renormalizability of the model can be traced back eventually to the absence of a metric.
There are no obvious counterterms to be written, a behaviour that could possibly
persist in four dimensions.

\section*{Acknowledgements}
We acknowledge the financial support from the RTN ENRAGE and the research projects 
FPA2007-66665 and SGR2009SGR502. This research is supported by the Consolidet CPAN
project. The work of J.A. was partially supported by Fondecyt 1060646.
We thank A. Andrianov and J. Russo for disccussions on the subject.

\end{document}